# Suspended Carbon Nanotube Quantum Wires with Two Gates

*Jien Cao, Qian Wang, Dunwei Wang and Hongjie Dai*[*]

Department of Chemistry and Laboratory for Advanced Materials, Stanford University,

Stanford, CA 94305

Suspended single-walled carbon nanotube devices comprised of high quality electrical contacts and two electrostatic gates per device are obtained. Compared to nanotubes pinned on substrates, the suspended devices exhibit little hysteresis related to environmental factors and manifest as cleaner Fabry-Perot interferometers or single electron transistors. The high-field saturation currents in suspended nanotubes related to optical phonon or zone boundary phonon scattering are significantly lower due to the lack of efficient heat sinking. The multiple gate design may also facilitate future investigation of electromechanical properties of nanotube quantum systems.

[*] Correspondence to hdai@stanford.edu

In recent years, successful fabrication of suspended single-walled carbon nanotube (SWNT) devices with electrical wiring has facilitated the investigation of the electrical, mechanical and electromechanical properties of these novel wires.[1-6] Here, we report a reliable method for the fabrication of suspended SWNT devices with improved metal-nanotube contacts and a useful double-gate design. These devices have led to the revelation of interesting electrical transport properties of nanotubes that are distinct from those pinned on substrates.

The structure of our suspended SWNT device is shown in Figure 1, consisting of a nanotube suspended over a trench by two pre-formed source (S) and drain (D) metal electrodes, a metal local-gate ($V_{GL}$) at the bottom of the trench and a global Si back-gate ($V_{GB}$). Device fabrication started with deposition of a 38 nm thick $Si_3N_4$ film on a p-type $SiO_2$/Si wafer ($SiO_2$ thickness = 500 nm). Electron-beam lithography (EBL) was used to define the local-gate pattern, dry-etching and wet-etching were used to open a window in the EBL resist and $Si_3N_4$ film and to further etch ~300 nm deep into the $SiO_2$ to form a trench with undercut (Figure 1a). Electron beam deposition of Pt/W (25 nm Pt; 5 nm W as sticking layer) followed by lift-off was then carried out to complete the formation of the trench with a Pt/W local gate electrode at the bottom. A second step EBL was performed to define the S/D pattern, followed by deposition of 5 nm W (as the sticking layer) and 25 nm Pt, and then liftoff. The region across the trench was written into a continuous stripe in the second EBL step. After metal deposition and liftoff, the stripe gave rise to the S/D electrodes separated by the trench. The S/D extended to the two edges of the trench and were electrically isolated from the local Pt gate due to undercutting in the sidewall of the trench. Finally, a third EBL step was carried out to pattern catalyst islands on top of the S/D. Chemical vapor deposition (CVD) was then used to grow SWNTs to bridge the electrode pairs on the wafer chip.[4,5] Low yield growth conditions were used (800 – 825 °C) to afford mostly single tubes bridging the S/D pairs, as confirmed by scanning electron microscope (SEM) after electrical measurements of the devices. The results presented here were all obtained from individual suspended SWNT devices.

Earlier, we described CVD growth of SWNTs across pre-formed Mo electrodes for electrical and electromechanical devices.[4] Mo was chosen as the S/D electrode material due to its compatibility with SWNT growth and the ability of dry-etching of Mo for electrode patterning. In the current work, our fabrication process does not involve dry-etching of metal and thus allows a wider choice of refractory metals. We succeeded in using refractory metals W and Pt/W (W as adhesive layer for Pt) as preformed electrodes and found that the typical resistance of our individual suspended SWNT devices was in the range of 10 – 50 kΩ (Figure 2 to 4). The contact resistance appeared higher than devices with nanotubes lying on substrates with Pd top-contacts[7,14], but lower compared to other suspended SWNT devices including those with preformed Mo electrodes[4] or top-contacted Au electrodes[3].

We observed several interesting properties for suspended SWNTs that are distinct from nanotubes pinned on substrates. The first is that the suspended SWNT devices are "hysteresis-free", for all the suspended devices we have measured (more than 100 devices): they show a small amount of hysteresis in current vs. gate ($I_{sd}$-$V_{GL}$) sweeps right after CVD growth, but the hysteresis completely disappears immediately after (in a few seconds) the devices are placed in Ar, dry air or vacuum (Figure 2a inset). This differs from nanotubes resting on SiO$_2$ substrates and is attributed to the lack of water molecules adsorbed on nanotubes in a dry environment, and that suspended nanotubes are free from intimate contact with water molecules adsorbed on SiO$_2$ substrate.[8]

The second interesting property of the suspended nanotube devices is that they appear to be 'cleaner' quantum systems than nanotubes on substrates. At low temperatures, the devices (length of suspension investigated here $L \sim$ 400-700 nm) exhibited characteristics corresponding to single quantum dots or interference resonators, without complex dots-in-series behavior often observed for tubes on substrates. A representative small band-gap semiconducting (SGS) SWNT[9] device is shown in Figure 2. The temperature dependence of the dip in the $I_{sd}$-$V_{GL}$ in Figure 2a shows activated transport. A small band gap of $E_g \sim$ 70 meV is obtained by fitting the conductance ($G$) at the dip to exp(-$E_g/2K_BT$). Transport through the p- and n-channels of the SWNT appear asymmetric, with higher conductance for the p-channel. At $T$=100 K, the subthreshold swings are $S \sim$ 150 mV/decade and 1.2 V/decade for the p- and n-subthreshold regions respectively (Figure 2a). At this temperature, the expected $S$ is [10] $S = \frac{1}{\alpha_{GL}} \cdot \frac{kT}{e} ln10 \approx 130 mV/decade$ if the Schottky barrier (SB) height is ~0, where $\alpha_{GL} \sim$ 0.14 is the local-gate efficiency factor (shift in Fermi level for a gate voltage change of $\Delta V_{GL}$ = 1V) as deduced later. Comparison of p- and n-channels' subthreshold swings to the theory predicted value suggests near zero and appreciable positive SBs[7,11] to the valence and conduction bands respectively between the SGS tube and the S/D electrodes. This is also consistent with the higher p-channel conductance than the n-channel.

A lock-in technique (503Hz, 50$\mu$V S/D bias) was used to perform $dI_{sd}/dV_{sd}$ measurements for the p-channel of our suspended-SWNT devices. When the local gate was set at a constant of $V_{GL} = -3.0$ V, $dI_{sd}/dV_{sd}$ vs. $V_{GB}$ and $V_{sd}$ exhibits an interference pattern with conductance peaks (~ 45 $\mu$S, $1.4e^2/h$) and valleys (~ 15 $\mu$S, $0.5e^2/h$) (Figure 2b). The conductance peaks at $V_{sd}$ ~ 2.8 mV corresponds to the length of the nanotube $L$ ~ 620 nm very well through $V_c = h v_F/2L$ ~ 1.67meV/$L(\mu m)$, suggesting that the p-channel of the SGS tube behaves as a well-defined Fabry-Perot interference wave guide or resonator.[12-14]

The n-channel of the device exhibits clear Coulomb blockade (CB) behavior (Figure 2c), as probed by fixing $V_{GL}$ = 3.5V and sweeping $V_{GB}$. The positive SBs to the conduction band give rises to two contact barriers that confine the nanotube quantum dot (QD). Clean periodic CB features corresponding to a single QD were observed with a charging energy of $E_c$ ~ 13 meV. Since the length of the nanotube is $L$ = 620 nm, a charging energy of ~13 meV is higher than $E_c$ ~ 5 meV/$L(\mu m)$ ~ 8 meV typical for nanotube QDs on SiO$_2$ substrates[15]. This is due to the low dielectric constant of the medium (air/vacuum) surrounding the suspended nanotube (compared to SiO$_2$ for tubes on substrate), which gives rise to a lower total capacitance and thus higher charging energy for the suspended nanotube. We find that $E_c$ ~ 8 meV/$L(\mu m)$ describes the charging energies of suspended SWNTs well. The local-gate efficiency is estimated to be $\alpha_{GL} = E_c/\Delta V_{GL} = 0.15$ where $\Delta V_{GL}$ ~ 87 mV is the period of Coulomb oscillation in Figure 2c. We also observed extra lines outside the CB diamonds (Figure 2c inset), corresponding to discrete excited quantum confined states along the length of the tube. The energy scale of the lines is ~ 2 meV, matching the discrete level spacing[15] of $\Delta E$ ~1 meV/$L(\mu m)$ in the nanotube.

We note that clean and homogeneous quantum interferometers and QDs are reliably observed with large numbers of our suspended-SWNT devices, at much higher frequency than tubes on substrates with similar lengths between S/D electrodes. This suggests few defects in our SWNTs at the sub-micron length scale and that the lack of substrate-nanotube interactions can prevent the break up of nanotubes into segments due to local chemical effects or mechanical strains.

A third interesting feature of our suspended nanotube devices is the two-gate configuration. We have measured the conductance of the sample versus both gates, as shown in Figure 3. Conductance features (e.g., peaks, valleys) appear as parallel lines in the 2-D plot, indicating the two gates are independent and additive. The slopes of the bright lines in $G$ vs. $V_{GL}$ and $V_{GB}$ can be used to deduce the ratio between the efficiencies of the two gates, $\alpha_{GL}/\alpha_{GB} \sim 30$. That is, the local-gate is $\sim 30$ times more efficient than the back-gate, or the local gate capacitance $C_{GL} \sim 30 C_{GB}$. The low efficiency of the back-gate can be understood due to screening of the local gate (Figure 1). Note that the local gate capacitance can be predicted by[16] $C_{GL} \approx \frac{2\pi\varepsilon\varepsilon_0 L}{ln(2h/r)} \approx 5 \times 10^{-18} F$, where $\varepsilon = 1$, $r \sim 1$ nm is the nanotube radius and $h = 300$ nm is the distance between the local gate and the nanotube (depth of trench). This suggests that $C_{GB} \sim C_{GL}/30 \sim 1.7 \times 10^{-19} F$, which is in reasonable agreement with $C_{GB} \sim 0.9 \times 10^{-19} F$ as measured from the CB data in the inset of Figure 2c. A potentially powerful aspect of the two-gate configuration is that under appropriate conditions, the local gate electrostatic coupling could become sufficiently strong to induce appreciable mechanical strain in the nanotube. The less effective global back-gate could then be used as a sweeping gate to characterize the electromechanical effects to the suspended nanotube quantum dot. We are currently pursuing this interesting possibility, especially for electromechanical measurements at low temperatures.

The fourth interesting result obtained with our suspended SWNT devices concerns with the high-bias transport properties of nanotubes when they are not resting on underlying substrates. This can shed light into the properties of nanotubes in their intrinsic states without significant roles played by the environment. For all of the suspended nanotubes that we have measured ($L \sim 400 – 700$nm, resistance $\sim 40 - 60$ k$\Omega$), we have consistently observed current saturation at the $I_{sat} \sim 8$ $\mu$A level, which is significantly lower than $\sim 15$-$20$ $\mu$A saturation currents for nanotubes (with similar resistance) lying on substrates. Since current saturation at high fields in SWNTs is caused by scattering of energetic ($\sim 0.2$ eV) optical or zone-boundary phonons,[17-19] the lower saturation current in a suspended nanotube can be understood by the lack of a proximal thermally conductive $SiO_2$ substrate for 'heat sinking'. Electrical heating is rapid in suspended nanotubes and the heat cannot be efficiently conducted away to the surrounding. This leads to increased acoustic phonon scattering, which is responsible for the observed current decrease under increasing bias (slight negative differential conductance feature in Figure 4).

In summary, a reliable method has been developed to obtain suspended SWNT devices with relatively good electrical contacts. Compared to nanotubes lying on substrates, the suspended devices exhibit little hysteresis behavior and manifest as well-defined single quantum dots or resonators. The high-field saturation current caused by optical phonon scattering in suspended nanotubes is low. The multiple gate design may facilitate future investigation of electromechanical properties of nanotubes.

This work was supported by MARCO MSD, a NSF-NIRT grant, SRC/AMD, DARPA/MTO, a Packard Fellowship, Sloan Research Fellowship and a Camille Dreyfus Teacher-Scholar Award.

**Figure Captions:**

**Figure 1.** (a) A schematic device structure. (b) SEM images of a device consisting of a single suspended SWNT and two gates (local gate $V_{GL}$ and global back-gate $V_{GB}$).

**Figure 2.** (a) $I_{sd}$-$V_{GL}$ (current vs. local gate) characteristics for a suspended small band-gap SWNT device (with back-gate grounded) recorded from room temperature to 1.5 K (curve with drastic oscillations) using $V_{sd}$=1mV bias. Inset: Hysteresis free double-sweep $I_{sd}$-$V_{GL}$ data recorded at room temperature. The two arrows correspond to back-and-forth gate sweeps. The $I_{sd}$-$V_{GL}$ curves recorded under the two sweep-directions overlap with each other. (b) A 2-D plot of conductance $dI_{sd}/dV_{sd}$ vs. $V_{GB}$ and bias $V_{sd}$ ($V_{GL}$ fixed at -3.0V) for the p-channel of the SWNT. The color scale bar for the conductance at the right of the graph is in unit of $e^2/h$. (c) A zoom-in of (a) in a smaller gate range $V_{GL}$ of 2 - 5V (bias=1mV) for the n-channel of the SWNT. Inset: A 2-D plot of $dI_{sd}/dV_{sd}$ vs. $V_{GB}$ and $V_{sd}$ ($V_{GL}$ set at 3.5 V). The two lines were drawn to highlight discrete electronic state due to quantum confinement along the tube length. Dark to bright colors correspond to conductance in the range of 0 to $0.2e^2/h$.

**Figure 3.** A 2-D plot of conductance $G$ vs. $V_{GL}$ and $V_{GB}$ with $V_{sd}$ = 1 mV. The bright color corresponds to high conductance of 1.4 $e^2/h$ and dark corresponds to zero conductance.

**Figure 4.** $I_{sd}$-$V_{sd}$ curve recorded for a suspended SWNT up to a high bias of $V_{sd}$ = 2 V. The dashed line was drawn to highlight the slight negative differential conductance behavior.

**References**

[1]  Tombler, T.W.; Zhou, CW.; Alexseyev, L.; Kong, J.; Dai, H.J.; Lei, L.; Jayanthi, C.S.; Tang, M.J.; Wu, S.Y. *Nature* **2000**, *405*, 769.

[2]  Walters, D.A.; Ericson, L.M.; Casavant, M.J.; Liu, J.; Colbert, D.T.; Smith, K.A.; Smalley, R.E., *Appl. Phys. Lett.* **1999**, *74*, 3803.

[3]  Nygard, J.; Cobden, D.H., *Appl. Phys. Lett.* **2001**, *79*, 4216.

[4]  Franklin, N.R.; Wang, Q.; Tombler, T.W.; Javey, A.; Shim, M.; Dai, H.J., *Appl. Phys. Lett.* **2002,** *81*, 913.

[5]  Cao J., Wang Q., Dai H.J., *Phys. Rev. Lett.* **2003,** *90*, 157601.

[6]  Minot, E.D.; Yaish, Y.; Sazonova, V.; Park, J.Y.; Brink, M.; McEuen, P.L., *Phys. Rev. Lett.* **2003,** *90*, 156401.

[7]  Javey, A.; Guo, J.; Wang, Q.; Lundstrom, M.; Dai, H.J., *Nature* **2003,** *424*, 654.

[8]  Kim, W.; Javey, A.; Vermesh, O.; Wang, Q.; Li, Y.M.; Dai, H.J., *Nano Lett.* **2003,** *3*, 193.

[9]  Zhou, C.W.; Kong, J.; Dai, H.J., *Phys. Rev. Lett.* **2000,** *84*, 5604.

[10] Sze, S.M., *Physics of semiconductor devices*; Wiley, New York, 1981.

[11] Heinze, S.; Tersoff, J.; Martel, R.; Derycke, V.; Appenzeller, J.; Avouris, P., *Phys. Rev. Lett.* **2002**, *89*, 6801.

[12] Liang, W.; Bockrath, M.; Bozovic, D.; Hafner, J.; Tinkham, M.; Park, H., *Nature* **2001**, *411*, 665.

[13] Kong, J.; Yenilmez, E.; Tombler, T. W.; Kim, W.; Liu, L.; Jayanthi, C. S.; Wu, S. Y.; Laughlin, R. B.; Dai, H. *Phys. Rev. Lett.* **2001**, *87*, 106801.

[14] Mann, D.; Javey, A.; Kong, J.; Wang, Q.; Dai, H.J., *Nano Lett.* **2003,** *3*, 1541.

[15] Nygard, J.; Cobden, D.H.; Bockrath, M.; McEuen, P.L.; Lindelof, P.E., *Appl. Phys. A* **1999,** *69*, 297.

[16] Ramo S., Whinnery J.R., Duzer T.V., *Fields and Waves in Communication Electronics*; Wiley, New York, 1994.

[17] Yao, Z.; Kane, C.L.; Dekker, C., *Phys. Rev. Lett.* **2000,** *84*, 2941.

[18] Javey A., Guo J., Paulsson M., Wang Q., Mann D., Lundstrom M., Dai H.J., *Phys. Rev. Lett.* **2004**, *92*, 106804.

[19] Park J.Y., Rosenblatt S., Yaish Y., Sazonova V., Ustunel H., Braig S., Arias T.A., Brouwer P.W., McEuen P.L., *Nano Lett.* **2004**, *4*, 517.

**Text for *Table of Contents***

Devices comprised of as-grown suspended single-walled carbon nanotube, high quality metal electrical contacts and two electrostatic gates per device are obtained for the first time. As shown in the graph, the length of the nanotube is fully suspended between two mesa metal contacts and free of any nanotube-substrate interaction. These suspended nanotube devices are ultra "clean" quantum systems and provide unique opportunity to investigate the intrinsic properties of nanotubes including transport behavior at high bias. In addition, the multiple gate design may also facilitate future work on electromechanical properties of suspended nanotube quantum systems.

**Keywords:**

Carbon nanotubes, Quantum dots, Electron Transport

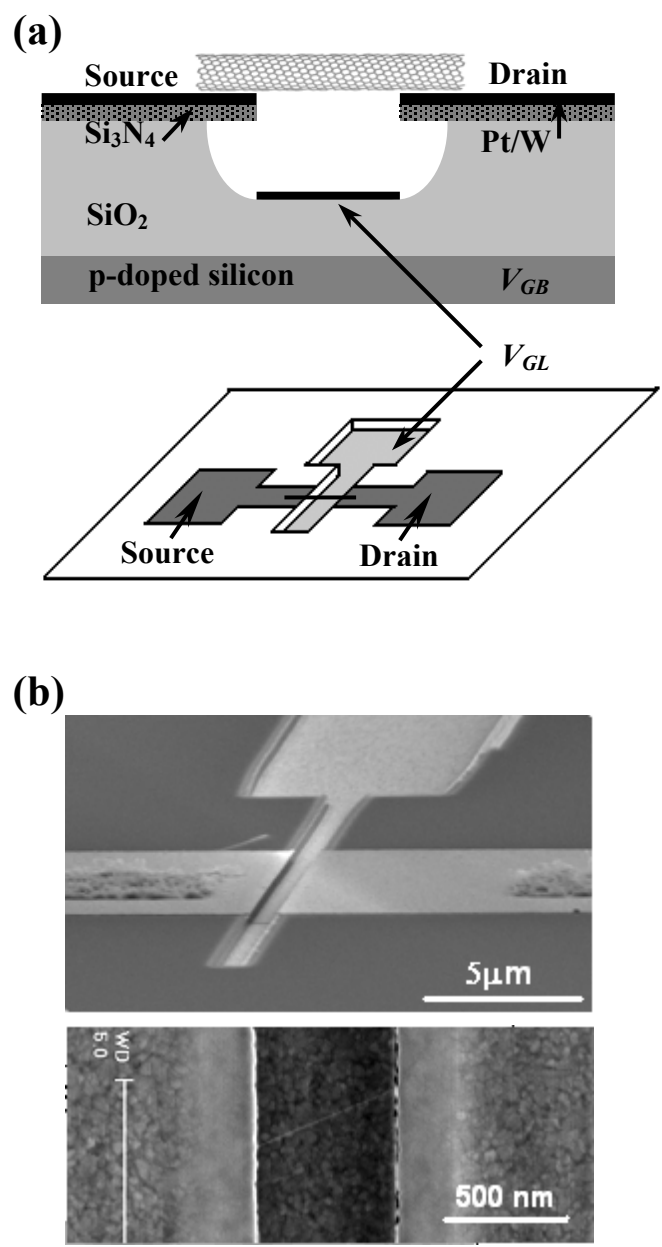

**Figure 1**

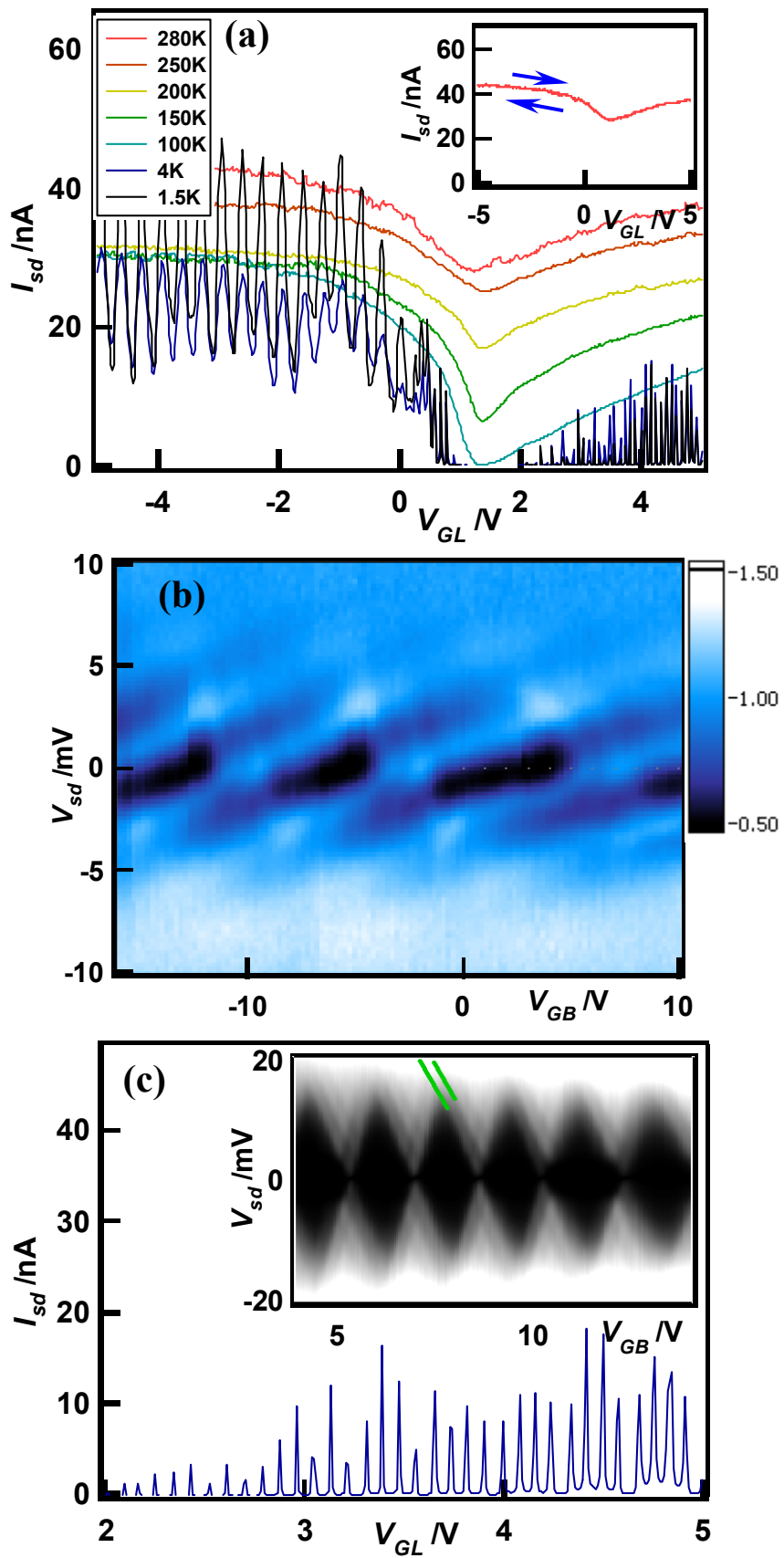

**Figure 2**

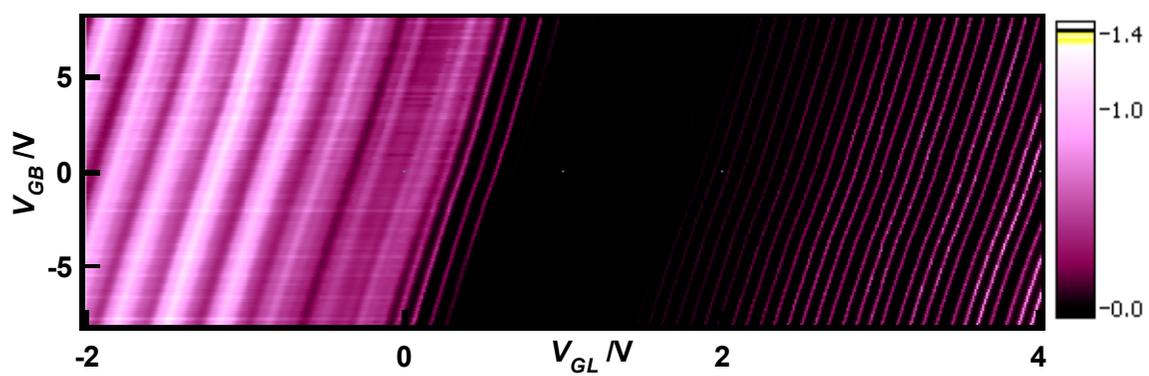

**Figure 3**

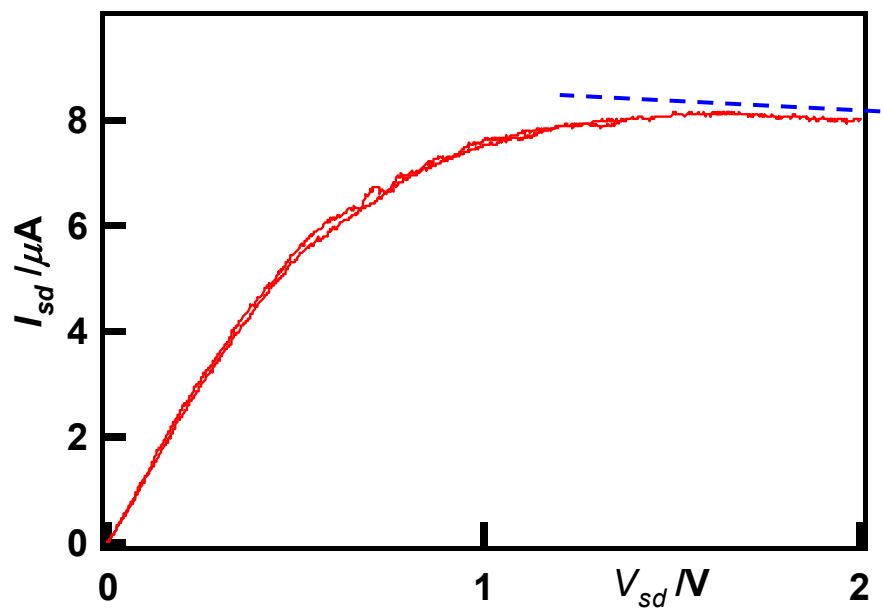

**Figure 4**